\documentclass[12pt]{article}
\usepackage{latexsym,amssymb,amsmath}
\begin{document}
\baselineskip=20pt

\newcommand{\la}{\langle}
\newcommand{\ra}{\rangle}
\newcommand{\psp}{\vspace{0.4cm}}
\newcommand{\pse}{\vspace{0.2cm}}
\newcommand{\ptl}{\partial}
\newcommand{\dlt}{\delta}
\newcommand{\sgm}{\sigma}
\newcommand{\al}{\alpha}
\newcommand{\be}{\beta}
\newcommand{\G}{\Gamma}
\newcommand{\gm}{\gamma}
\newcommand{\vs}{\varsigma}
\newcommand{\Lmd}{\Lambda}
\newcommand{\lmd}{\lambda}
\newcommand{\td}{\tilde}
\newcommand{\vf}{\varphi}
\newcommand{\yt}{Y^{\nu}}
\newcommand{\wt}{\mbox{wt}\:}
\newcommand{\rd}{\mbox{Res}}
\newcommand{\ad}{\mbox{ad}}
\newcommand{\stl}{\stackrel}
\newcommand{\ol}{\overline}
\newcommand{\ul}{\underline}
\newcommand{\es}{\epsilon}
\newcommand{\dmd}{\diamond}
\newcommand{\clt}{\clubsuit}
\newcommand{\vt}{\vartheta}
\newcommand{\ves}{\varepsilon}
\newcommand{\dg}{\dagger}
\newcommand{\tr}{\mbox{Tr}}
\newcommand{\ga}{{\cal G}({\cal A})}
\newcommand{\hga}{\hat{\cal G}({\cal A})}
\newcommand{\Edo}{\mbox{End}\:}
\newcommand{\for}{\mbox{for}}
\newcommand{\kn}{\mbox{ker}}
\newcommand{\Dlt}{\Delta}
\newcommand{\rad}{\mbox{Rad}}
\newcommand{\rta}{\rightarrow}
\newcommand{\mbb}{\mathbb}
\newcommand{\lra}{\Longrightarrow}
\newcommand{\X}{{\cal X}}
\newcommand{\Y}{{\cal Y}}
\newcommand{\Z}{{\cal Z}}
\newcommand{\U}{{\cal U}}
\newcommand{\V}{{\cal V}}
\newcommand{\W}{{\cal W}}
\newcommand{\sech}{\mbox{sech}\:}
\newcommand{\csch}{\mbox{csch}\:}
\newcommand{\sn}{\mbox{sn}\:}
\newcommand{\cn}{\mbox{cn}\:}
\newcommand{\dn}{\mbox{dn}\:}

\begin{center}{\Large \bf Quadratic-Argument  Approach to}
\end{center}
\begin{center}{\Large \bf the Davey-Stewartson  Equations}
\end{center}

\vspace{0.2cm}

\begin{center}{\large Xiaoping Xu}\end{center}
\begin{center}{Institute of Mathematics, Academy of Mathematics \& System Sciences}\end{center}
\begin{center}{Chinese Academy of Sciences, Beijing 100190, P.R. China}
\footnote{Research supported
 by China NSF 10871193}\end{center}

\vspace{0.6cm}

 \begin{center}{\Large\bf Abstract}\end{center}

\vspace{1cm} {\small  The Davey-Stewartson  equations are used to
describe the long time evolution of a three-dimensional packets of
surface waves. Assuming that the argument functions are quadratic in
spacial variables, we find in this paper various
 exact solutions  modulo the most known symmetry transformations for the Davey-Stewartson  equations.}
\psp


\section{Introduction}

 The Davey and Stewartson [6] (1974) used the method of multiple scales
 to derive the following system of nonlinear partial differential
 equations
$$2iu_t+ \es_1u_{xx}+u_{yy}-2\es_2|u|^2u-2uv=0,\eqno(1.1)$$
$$v_{xx}-\es_1(v_{yy}+2(|u|^2)_{xx})=0\eqno(1.2)$$
that describe the long time evolution of a three-dimensional packets
of surface waves, where $u$ is a complex-valued function, $v$ is a
real valued function and $\es_1,\es_2=\pm 1$. The equations are
called  the {\it Davey-Stewartson I equations} if $\es_1=1$, and the
{\it Davey-Stewartson II equations} when $\es_1=-1$. They were used
to study the stability of the uniform Stokes wave train with respect
to small disturbance. The soliton solutions of the Davey-Stewartson
equations were first studied by Anker and Freeman [2] (1978). Kirby
and Dalrymple [7] (1983) obtained oblique envelope solutions of the
equations in intermediate water depth. Omote [13] (1988) found
infinite-dimensional symmetry algebras and an infinite number of
conserved quantities for the equations.

Arkadiev,  Pogrebkov and  Polivanov [3] (1989) studied the solutions
of the Davey-Stewartson II equations whose singularities form closed
lines with string-like behavior. They [4] (1989) also applied the
inverse sacttering transform method to the Davey-Stewartson II
equations. We refer [1] for more information on the Davey-Stewartson
equations in terms of inverse scattering and integrable systems.
Gilson and Nimmo [8] (1991) found dromion solutions and Malanyuk
[11, 12] (1991, 1994) obtained finite-gap solutions of the
equations. van de Linden [15] (1992) studied the solutions under a
certain boundary condition. Clarkson and Hood [5] (1994) obtained
certain symmetry reductions of the equations to ordinary
differential equations with no intervening steps and provided new
exact solutions which are not obtainable by the Lie group approach.
Guil and Manas [9] (1995) found certains solutions of the
Davey-Stewartson I equations by deforming dromion. Manas and Santini
[10] (1997) studied a large class of solutions of the
Davey-Stewartson II equations by a Wronskian scheme. It is obvious
that the some of above solutions are equivalent to each other under
the known symmetric transformations. It is time to study solutions
of the Davey-Stewartson equations modulo the known symmetric
transformations.

We observed in [17] that the argument functions of all the solutions
of the two-dimensional cubic nonlinear  Schr\"{o}dinger equation in
[15] are quadratic in spacial variables. This motivated us [17] to
introduce a quadratic-argument approach to study exact solutions of
 the two-dimensional cubic nonlinear Schr\"{o}dinger equation and the
coupled two-dimensional cubic nonlinear Schr\"{o}dinger equations
modulo the known symmetry transformations. Indeed, our solution sets
are most complete among the ones whose argument functions are
quadratic in spacial variables. In this paper, we use the
quadratic-argument approach to study exact solutions of  the
Davey-Stewartson equations modulo the most known symmetry
transformations.

For convenience, we always assume that all the involved partial
derivatives of related functions always exist and we can change
orders of taking partial derivatives. We also use prime $'$ to
denote the derivative of any one-variable function. The known
symmetry transformations we are concerned with are:
$${\cal
T}_1(u)=e^{-(\es_1\al'x+\be'y+\gm)i}\xi(t,x+\al,y+\be),\eqno(1.3)$$
$${\cal
T}_1(v)=v(t,x+\al,y+\be)+\es_1{\al'}'x+{\be'}'y-\frac{\es_1(\al')^2+(\be')^2}{2}+\gm';\eqno(1.4)$$
$${\cal T}_2(u)=b^{-1}u(b^2t,bx,by),\qquad {\cal
T}_2(v)=b^{-2}v(b^2t,bx,by);\eqno(1.5)$$ where $\al,\be,\gm$ are
functions of $t$ and $b$ is a nonzero real constant.  The above
transformations transform one solution of (1.1) and (1.2) into
another solution. In particular, applying the transformation ${\cal
T}_1$ to any known solution  would yield solutions with three
additional parameter functions.

\section{Exact Solutions}

 Write
$$u=\xi(t,x,y)e^{i\phi(t,x,y)},\eqno(2.1)$$
where $\xi$ and $\phi$ are real functions in $t,x,y$. Note
$$u_t=(\xi_t+i\xi\phi_t)e^{i\phi},\qquad
u_x=(\xi_x+i\xi\phi_x)e^{i\phi},\qquad
u_y=(\xi_y+i\xi\phi_y)e^{i\phi},\eqno(2.2)$$
$$u_{xx}=(\xi_{xx}-\xi\phi_x^2+i(2\xi_x\phi_x
+\xi\phi_{xx}))e^{i\phi},\;\;
u_{yy}=(\xi_{yy}-\xi\phi_y^2+i(2\xi_y\phi_y
+\xi\phi_{yy}))e^{i\phi}.\eqno(2.3)$$ Then (1.1) is equivalent to
\begin{eqnarray*}\hspace{2cm}& &2i\xi_t-2\xi\phi_t+\es_1(\xi_{xx}-\xi\phi_x^2+i(2\xi_x\phi_x
+\xi\phi_{xx}))\\ &&+\xi_{yy}-\xi\phi_y^2+i(2\xi_y\phi_y
+\xi\phi_{yy})-2\es_2\xi^3-2\xi
v=0,\hspace{3.7cm}(2.4)\end{eqnarray*} equivalently,
$$2\xi_t+2(\es_1\xi_x\phi_x+\xi_y\phi_y)
+\xi(\es_1\phi_{xx}+\phi_{yy})=0,\eqno(2.5)$$
$$\xi(2\phi_t+\es_1\phi_x^2+\phi_y^2)-\es_1\xi_{xx}-\xi_{yy}+2\es_2\xi^3+2\xi
v=0.\eqno(2.6)$$ Moreover, (1.2) becomes
$$v_{xx}-\es_1(v_{yy}+2(\xi^2)_{xx})=0.\eqno(2.7)$$

In order to solve the above system of partial differential
equations, we assume in this section that
$$\phi=\al'x^2+\be'y^2\eqno(2.8)$$ for
some functions $\al$ and $\be$ of $t$. Then (2.5) becomes
$$\xi_t+2(\es_1\al'x\xi_x+\be' y\xi_y)
+(\es_1\al'+\be')\xi=0.\eqno(2.9)$$ Thus
$$\xi=e^{-\es_1\al-\be}\vt(\varpi_1,\varpi_2),\qquad
\varpi_1=e^{-2\es_1\al}x,\;\varpi_2=e^{-2\be}y,\eqno(2.10)$$ where
$\vt$ is some two-variable function in $\varpi_1$ and $\varpi_2$.
Moreover, (2.6) is equivalent to
\begin{eqnarray*}\hspace{2cm}& &2(({\al'}'+2\es_1(\al')^2)x^2+({\be'}'+2(\be')^2)y^2)\vt-\es_1e^{-4\es_1\al}
\vt_{\varpi_1\varpi_1}\\ &
&-e^{-4\be}\vt_{\varpi_2\varpi_2}+2\es_2e^{-2\es_1\al-2\be}\vt^3+2\vt
v=0.\hspace{5cm}(2.11)\end{eqnarray*} \pse

{\it Case 1}. $\vt=a\varpi_1+b\varpi_2+c$ for $a,b,c\in\mbb{R}$.

In this case, The above equation is equivalent to
$$({\al'}'+2\es_1(\al')^2)x^2+({\be'}'+2(\be')^2)y^2+\es_2e^{-2\es_1\al-2\be}(ae^{-2\es_1\al}x+be^{-2\be}y+c)^2+v=0.\eqno(2.12)$$
So
\begin{eqnarray*}v&=&-[{\al'}'+2\es_1(\al')^2+\es_2a^2e^{-6\es_1\al-2\be}]x^2
-[{\be'}'+2(\be')^2+\es_2b^2e^{-2\es_1\al-6\be}]y^2\\
&
&-2ab\es_2e^{-4\es_1\al-4\be}xy-\es_2ce^{-2\es_1\al-2\be}(2ae^{-2\es_1\al}x+2be^{-2\be}y+c).
\hspace{3cm}(2.13)\end{eqnarray*} Moreover, (2.7) is equivalent to
$${\al'}'+2\es_1(\al')^2+\es_2a^2e^{-6\es_1\al-2\be}=
\es_1[{\be'}'+2(\be')^2+\es_2b^2e^{-2\es_1\al-6\be}].\eqno(2.14)$$
To solve the above equation, we write
$$\al=\Im+\es_1\be.\eqno(2.15)$$
Then the above equation becomes
$${\Im'}'+2\es_1(\Im')^2+4\Im'\be'+\es_2(a^2e^{-6\es_1\Im}-\es_1b^2e^{-2\es_1\Im})e^{-8\be}=0.\eqno(2.16)$$

Assuming $ab\neq 0$ and $\es_1=1$, we have the following simple
solution
$$\Im=\frac{1}{2}(\ln |a|-\ln|b|),\eqno(2.17)$$
 where $\be$ is any function of
$t$. Suppose $a=b=0$. Then we take
$$\be'=-\frac{{\Im'}'}{4\Im'}-\frac{\es_1\Im'}{2}\lra\be=-\frac{1}{4}\ln \Im'-\frac{\es_1\Im}{2}\eqno(2.18)$$
modulo the transformation in (1.5), where $\Im$ is arbitrary
increasing function of $t$. Then
$$\al=(\Im')^{-\es_1/4}e^{\Im/2},\qquad
\be=(\Im')^{-/4}e^{-\es_1\Im/2}.\eqno(2.19)$$ By (2.8), (2.10) and
(2.13), we have:\psp

{\bf Theorem 2.1}. {\it  Let $\Im$ be any increasing function of $t$
and let $a,b,c\in\mbb{R}$ with $ab\neq 0$. Suppose that $\be$ is any
function of $t$. We have the following solutions of the
Davey-Stewartson equations (1.1) and (1.2):
\begin{eqnarray*}\hspace{1.3cm}u&=&c\exp(-\es_1(\Im')^{-\es_1/4}e^{\Im/2}-(\Im')^{-/4}e^{-\es_1\Im/2})\\ & &\times\exp
\frac{(2(\Im')^2-\es_1{\Im'}')x^2-(2\es_1(\Im')^2+{\Im'}')y^2}{4\Im'}i,
\hspace{4cm}(2.20)\end{eqnarray*}
\begin{eqnarray*}v&=&
\frac{\es_1({{\Im'}'}'-({\Im'}')^2)-2(\Im')^2{\Im'}'}{4(\Im')^2}x^2+
\frac{{{\Im'}'}'-({\Im'}')^2+2\es_1(\Im')^2{\Im'}'}{4(\Im')^2}y^2\\
&
&-\es_2c^2\exp(-2\es_1(\Im')^{-\es_1/4}e^{\Im/2}-2(\Im')^{-/4}e^{-\es_1\Im/2})
; \hspace{5cm}(2.21)\end{eqnarray*}if $\es_1=1$,
$$u=e^{\be'(x^2+y^2)i-(2\be+(\ln |a|-\ln|b|)/2)}(ae^{\ln|b|-\ln
|a|-2\be}x+be^{-2\be}y+c),\eqno(2.22)$$
\begin{eqnarray*}\hspace{1.3cm}v&=&-[{\be'}'+2(\be')^2+\es_2|a|^{-1}|b|^3e^{-8\be}](x^2+y^2)
-2a^{-1}b^3\es_2e^{-8\be}xy\\&
&-\es_2ce^{-4\be}(2a^{-1}b^2e^{-2\be}x+2|a|^{-1}b|b|e^{-2\be}y+c).
\hspace{4.3cm}(2.23)\end{eqnarray*}} \pse

{\bf Remark 2.2}. Applying the symmetry transformation ${\cal T}_1$
in (1.3) and (1.4) to the above solutions, we get the following
solutions with additional three parameter functions
$\al_1,\be_1,\gm_1$ of $t$:
\begin{eqnarray*}u&=&c\exp
\left(\frac{(2(\Im')^2-\es_1{\Im'}')(x+\al_1)^2-(2\es_1(\Im')^2+{\Im'}')(y+\be_1)^2}{4\Im'}
-\es_1\al'_1x-\be'_1y-\gm_1\right)i\\ & &\times
\exp(-\es_1(\Im')^{-\es_1/4}e^{\Im/2}-(\Im')^{-/4}e^{-\es_1\Im/2}),
, \hspace{6cm}(2.24)\end{eqnarray*}
\begin{eqnarray*}v=
\frac{\es_1({{\Im'}'}'-({\Im'}')^2)-2(\Im')^2{\Im'}'}{4(\Im')^2}(x+\al_1)^2+
\frac{{{\Im'}'}'-({\Im'}')^2+2\es_1(\Im')^2{\Im'}'}{4(\Im')^2}(y+\be_1)^2+\es_1{\al_1'}'x\\
+{\be_1'}'y-\es_2c^2\exp(-2\es_1(\Im')^{-\es_1/4}e^{\Im/2}-2(\Im')^{-/4}e^{-\es_1\Im/2})-\frac{\es_1(\al_1')^2+(\be_1')^2}{2}
+\gm_1' ; \hspace{1cm}(2.25)\end{eqnarray*}if $\es_1=1$,
\begin{eqnarray*}\hspace{1.2cm}u&=&e^{[\be'(x^2+y^2)-\es_1\al'_1x+\be'_1y+\gm_1]i-(2\be+(\ln
|a|-\ln|b|)/2)}\\ & &\times (ae^{\ln|b|-\ln
|a|-2\be}(x+\al_1)+be^{-2\be}(y+\be_1)+c),\hspace{4cm}(2.26)\end{eqnarray*}
\begin{eqnarray*}v&=&-[{\be'}'+2(\be')^2+\es_2|a|^{-1}|b|^3e^{-8\be}]((x+\al_1)^2+(y+\be_1)^2)
+\es_1{\al_1'}'x+{\be_1'}'y\\&
&-\es_2ce^{-4\be}(2a^{-1}b^2e^{-2\be}(x+\al_1)+2|a|^{-1}b|b|e^{-2\be}(y+\be_1)+c)\\
&
&-2a^{-1}b^3\es_2e^{-8\be}(x+\al_1)(y+\be_1)-\frac{\es_1(\al_1')^2+(\be_1')^2}{2}+\gm_1'
. \hspace{3.8cm}(2.27)\end{eqnarray*}\pse

{\it Case 2}. $\es_1\al=\be$.\psp

We set
$$\zeta_1(s)=\sinh s,\;\;\eta_1(s)=\cosh s,\;\;\zeta_{-1}(s)=\sin
s,\;\;\eta_{-1}(s)=\cos s.\eqno(2.28)$$ Let $\ell$ and $\ell_1$ be
two fixed real constants. We denote
$$\varpi=\zeta_{\es_1}(\ell)\varpi_1+\eta_{\es_1}(\ell)\varpi_2+\ell_1.\eqno(2.29)$$
Moreover, we assume
$$\vt=\nu(\varpi).\eqno(2.26)$$
Observe
$$(\ptl_x^2-\es_1\ptl_y^2)(\nu(\varpi))=-\es_1{\nu'}'(\varpi)\eqno(2.30)$$
and
$$(\es_1\ptl_x^2+\ptl_y^2)(\nu(\varpi))=(\eta_{\es_1}(2\ell)){\nu'}'(\varpi).\eqno(2.31)$$
By (2.7), we further assume
$$v=\gm(\es_1x^2+y^2)+e^{-4\be}(c+\zeta_{\es_1}^2(\ell)\nu^2)\eqno(2.32)$$ for a function $\gm$ of
$t$ and a real constant $c$. So (2.7) naturally holds.

 Now (2.11) becomes
$$2({\be'}'+2(\be')^2+\gm)(\es_1x^2+y^2)\nu
+e^{-4\be}(c\nu-\eta_{\es_1}(2\ell){\nu'}'+2(\es_2+\zeta_{\es_1}^2(\ell))\nu^3)=0.\eqno(2.33)$$
For simplicity, we take
$$\gm=-{\be'}'-2(\be')^2.\eqno(2.34)$$
So (2.29) is equivalent to
$$c\nu-\eta_{\es_1}(2\ell){\nu'}'+2(\es_2+\zeta_{\es_1}^2(\ell))\nu^3=0.\eqno(2.35)$$
First we have simple solution
$$\nu=\frac{1}{\varpi}\sqrt{\frac{\eta_{\es_1}(2\ell)}{\es_2+\zeta_{\es_1}^2(\ell)}},\qquad c=0\eqno(2.36)$$

Recall
$${(\tan s)'}'=2(\tan^3 s+\tan
s), \qquad{(\sec s)'}'=2\sec^3 s-\sec s,\eqno(2.37)$$
$${(\coth s)'}'=2(\coth^3s-\coth s),\qquad
{(\csch s)'}'=2\csch^3 s+\csch s.\eqno(2.38)$$ Denote Jacobi
elliptic functions
$$\sn s=\sn(s|m),\qquad\cn s=\cn(s|m),\qquad\dn s=\dn(s|m),\eqno(2.39)$$
where $m$ is the elliptic modulus (e.g., cf. [16]). Then
$${(\sn s)'}'=2m^2\sn^3s-(1+m^2)\sn s,\eqno(2.40)$$
$${(\cn s)'}'=-2m^2\cn^2s+(2m^2-1)\cn s,\eqno(2.41)$$
$${(\dn s)'}'=-2\dn^3s+(2-m^2)\dn s.\eqno(2.42)$$
Moreover,
$$\lim_{m\rta 1}\sn s=\tanh s,\qquad \lim_{m\rta 1}\cn s=
\lim_{m\rta 1}\dn s=\sech s.\eqno(2.43)$$

Comparing (2.35) with the equations in (2.37), (2.38) and
(2.39)-(2.42),
 we have the following
solutions:
$$\nu=\sqrt{\frac{\eta_{\es_1}(2\ell)}{\es_2+\zeta_{\es_1}^2(\ell)}}\:\tan \varpi,\qquad
c=2\eta_{\es_1}(2\ell);\eqno(2.44)$$
$$\nu=\sqrt{\frac{\eta_{\es_1}(2\ell)}{\es_2+\zeta_{\es_1}^2(\ell)}}\:\sec \varpi,\qquad
 c=-\eta_{\es_1}(2\ell);\eqno(2.45)$$
$$\nu=\sqrt{\frac{\eta_{\es_1}(2\ell)}{\es_2+\zeta_{\es_1}^2(\ell)}}\:\coth \varpi,\qquad
 c=-2\eta_{\es_1}(2\ell);\eqno(2.46)$$
$$\nu=\sqrt{\frac{\eta_{\es_1}(2\ell)}{\es_2+\zeta_{\es_1}^2(\ell)}}\:\csch\varpi,
\qquad c=\eta_{\es_1}(2\ell);\eqno(2.47)$$
$$\nu=m\sqrt{\frac{\eta_{\es_1}(2\ell)}{\es_2+\zeta_{\es_1}^2(\ell)}}\:\sn \varpi,
\qquad c=-(1+m^2)\eta_{\es_1}(2\ell);\eqno(2.48)$$
$$\nu=m\sqrt{\frac{-\eta_{\es_1}(2\ell)}{\es_2+\zeta_{\es_1}^2(\ell)}}\:\cn \varpi,\qquad
c=(2m^2-1)\eta_{\es_1}(2\ell),\eqno(2.49)$$
$$\nu=\sqrt{\frac{-\eta_{\es_1}(2\ell)}{\es_2+\zeta_{\es_1}^2(\ell)}}\:\dn \varpi,\qquad
 c=(2-m^2)\eta_{\es_1}(2\ell).\eqno(2.50)$$
By (2.8), (2.10) and (2.32), we have: \psp

{\bf Theorem 2.3}. {\it Let $\ell,\ell_1,m\in\mbb{R}$.  Suppose that
$\be$ is any function of $t$. Take the notations in (2.28). The
followings are solutions of the Davey-Stewartson equations (1.1) and
(1.2) (where the solution exists only when the expression makes
sense as real function): }
$$u=\frac{e^{-2\be+(\es_1x^2+y^2)\be'i}}{e^{-2\be}(\zeta_{\es_1}(\ell)x+\eta_{\es_1}(\ell)y)+\ell_1}
\sqrt{\frac{\eta_{\es_1}(2\ell)}{\es_2+\zeta_{\es_1}^2(\ell)}},\eqno(2.51)$$
$$
v=-({\be'}'+2(\be')^2)(\es_1x^2+y^2)+\frac{e^{-4\be}\zeta_{\es_1}^2(\ell)\eta_{\es_1}(2\ell)}{
(\es_2+\zeta_{\es_1}^2(\ell))(e^{-2\be}(\zeta_{\es_1}(\ell)x+\eta_{\es_1}(\ell)y)+\ell_1)^2}
; \eqno(2.52)$$
$$u=\sqrt{\frac{\eta_{\es_1}(2\ell)}{\es_2+\zeta_{\es_1}^2(\ell)}}\:e^{-2\be+(\es_1x^2+y^2)\be'i}
\tan(e^{-2\be}(\zeta_{\es_1}(\ell)x+\eta_{\es_1}(\ell)y)+\ell_1),\eqno(2.53)$$
\begin{eqnarray*}v&=&e^{-4\be}\left(\frac{\zeta_{\es_1}^2(\ell)\eta_{\es_1}(2\ell)}{
\es_2+\zeta_{\es_1}^2(\ell)}\tan^2(e^{-2\be}(\zeta_{\es_1}(\ell)x+\eta_{\es_1}(\ell)y)+\ell_1)+2\eta_{\es_1}(2\ell)\right)
\\ &
&-({\be'}'+2(\be')^2)(\es_1x^2+y^2);\hspace{8.9cm}(2.54)\end{eqnarray*}
$$u=\sqrt{\frac{\eta_{\es_1}(2\ell)}{\es_2+\zeta_{\es_1}^2(\ell)}}\:e^{-2\be+(\es_1x^2+y^2)\be'i}
\sec(e^{-2\be}(\zeta_{\es_1}(\ell)x+\eta_{\es_1}(\ell)y)+\ell_1),\eqno(2.55)$$
\begin{eqnarray*}v&=&e^{-4\be}\left(\frac{\zeta_{\es_1}^2(\ell)\eta_{\es_1}(2\ell)}{
\es_2+\zeta_{\es_1}^2(\ell)}\sec^2(e^{-2\be}(\zeta_{\es_1}(\ell)x+\eta_{\es_1}(\ell)y)+\ell_1)-\eta_{\es_1}(2\ell)\right)
\\ &
&-({\be'}'+2(\be')^2)(\es_1x^2+y^2);\hspace{8.9cm}(2.56)\end{eqnarray*}
$$u=\sqrt{\frac{\eta_{\es_1}(2\ell)}{\es_2+\zeta_{\es_1}^2(\ell)}}\:e^{-2\be+(\es_1x^2+y^2)\be'i}
\coth(e^{-2\be}(\zeta_{\es_1}(\ell)x+\eta_{\es_1}(\ell)y)+\ell_1),\eqno(2.57)$$
\begin{eqnarray*}v&=&e^{-4\be}\left(\frac{\zeta_{\es_1}^2(\ell)\eta_{\es_1}(2\ell)}{
\es_2+\zeta_{\es_1}^2(\ell)}\coth^2(e^{-2\be}(\zeta_{\es_1}(\ell)x+\eta_{\es_1}(\ell)y)+\ell_1)-2\eta_{\es_1}(2\ell)\right)
\\ &
&-({\be'}'+2(\be')^2)(\es_1x^2+y^2);\hspace{8.9cm}(2.58)\end{eqnarray*}
$$u=\sqrt{\frac{\eta_{\es_1}(2\ell)}{\es_2+\zeta_{\es_1}^2(\ell)}}\:e^{-2\be+(\es_1x^2+y^2)\be'i}
\csch(e^{-2\be}(\zeta_{\es_1}(\ell)x+\eta_{\es_1}(\ell)y)+\ell_1),\eqno(2.59)$$
\begin{eqnarray*}v&=&e^{-4\be}\left(\frac{\zeta_{\es_1}^2(\ell)\eta_{\es_1}(2\ell)}{
\es_2+\zeta_{\es_1}^2(\ell)}\csch^2(e^{-2\be}(\zeta_{\es_1}(\ell)x+\eta_{\es_1}(\ell)y)+\ell_1)+\eta_{\es_1}(2\ell)\right)
\\ &
&-({\be'}'+2(\be')^2)(\es_1x^2+y^2);\hspace{8.9cm}(2.60)\end{eqnarray*}
$$u=m\sqrt{\frac{\eta_{\es_1}(2\ell)}{\es_2+\zeta_{\es_1}^2(\ell)}}\:e^{-2\be+(\es_1x^2+y^2)\be'i}
\sn(e^{-2\be}(\zeta_{\es_1}(\ell)x+\eta_{\es_1}(\ell)y)+\ell_1),\eqno(2.61)$$
\begin{eqnarray*}v&=&e^{-4\be}\left(\frac{m^2\zeta_{\es_1}^2(\ell)\eta_{\es_1}(2\ell)}{
\es_2+\zeta_{\es_1}^2(\ell)}\sn^2(e^{-2\be}(\zeta_{\es_1}(\ell)x+\eta_{\es_1}(\ell)y)+\ell_1)-(1+m^2)\eta_{\es_1}(2\ell)\right)
\\ &
&-({\be'}'+2(\be')^2)(\es_1x^2+y^2);\hspace{8.9cm}(2.62)\end{eqnarray*}
$$u=m\sqrt{\frac{-\eta_{\es_1}(2\ell)}{\es_2+\zeta_{\es_1}^2(\ell)}}\:e^{-2\be+(\es_1x^2+y^2)\be'i}
\cn(e^{-2\be}(\zeta_{\es_1}(\ell)x+\eta_{\es_1}(\ell)y)+\ell_1),\eqno(2.53)$$
\begin{eqnarray*}v&=&e^{-4\be}\left((2m^2-1)\eta_{\es_1}(2\ell)-\frac{m^2\zeta_{\es_1}^2(\ell)\eta_{\es_1}(2\ell)}{
\es_2+\zeta_{\es_1}^2(\ell)}\cn^2(e^{-2\be}(\zeta_{\es_1}(\ell)x+\eta_{\es_1}(\ell)y)+\ell_1)\right)
\\ &
&-({\be'}'+2(\be')^2)(\es_1x^2+y^2);\hspace{8.9cm}(2.64)\end{eqnarray*}
$$u=\sqrt{\frac{-\eta_{\es_1}(2\ell)}{\es_2+\zeta_{\es_1}^2(\ell)}}\:e^{-2\be+(\es_1x^2+y^2)\be'i}
\dn(e^{-2\be}(\zeta_{\es_1}(\ell)x+\eta_{\es_1}(\ell)y)+\ell_1),\eqno(2.65)$$
\begin{eqnarray*}v&=&e^{-4\be}\left((2-m^2)\eta_{\es_1}(2\ell)-\frac{m^2\zeta_{\es_1}^2(\ell)\eta_{\es_1}(2\ell)}{
\es_2+\zeta_{\es_1}^2(\ell)}\dn^2(e^{-2\be}(\zeta_{\es_1}(\ell)x+\eta_{\es_1}(\ell)y)+\ell_1)\right)
\\ &
&-({\be'}'+2(\be')^2)(\es_1x^2+y^2);\hspace{8.9cm}(2.66)\end{eqnarray*}
\pse

{\bf Remark 2.4}. Applying the symmetry transformation ${\cal T}_1$
in (1.3) and (1.4) to the above solutions, we can obtain solutions
with additional three parameter functions. For instance, we get the
following solutions with additional three parameter functions
$\al_1,\be_1,\gm_1$ of $t$ from the above first two solutions:
$$u=\frac{e^{-2\be+[(\es_1(x+\al_1)^2+(y+\be_1)^2)\be'-\es_1\al_1'x-\be_1'y-\gm_1]i}}{e^{-2\be}(\zeta_{\es_1}(\ell)(x+\al_1)+
\eta_{\es_1}(\ell)(y+\be_1))+\ell_1}
\sqrt{\frac{\eta_{\es_1}(2\ell)}{\es_2+\zeta_{\es_1}^2(\ell)}},\eqno(2.67)$$
\begin{eqnarray*}
v&=&-({\be'}'+2(\be')^2)(\es_1(x+\al_1)^2+(y+\be_1)^2)+\es_1{\al_1'}'x+{\be_1'}'y-\frac{\es_1(\al_1')^2+(\be_1')^2}{2}
+\gm_1\\ & &
+\frac{e^{-4\be}\zeta_{\es_1}^2(\ell)\eta_{\es_1}(2\ell)}{
(\es_2+\zeta_{\es_1}^2(\ell))(e^{-2\be}(\zeta_{\es_1}(\ell)(x+\al_1)+\eta_{\es_1}(\ell)(y+\be_1))+\ell_1)^2}
; \hspace{3.1cm}(2.68)\end{eqnarray*}
\begin{eqnarray*}\hspace{1cm}u&=&\sqrt{\frac{\eta_{\es_1}(2\ell)}{\es_2+\zeta_{\es_1}^2(\ell)}}\:
e^{-2\be+[(\es_1(x+\al_1)^2+(y+\be_1)^2)\be'-\es_1\al_1'x-\be_1'y-\gm_1]i}\\&
&\times
\tan(e^{-2\be}(\zeta_{\es_1}(\ell)(x+\al_1)+\eta_{\es_1}(\ell)(y+\be_1))+\ell_1),\hspace{3.8cm}(2.69)\end{eqnarray*}
\begin{eqnarray*}v=e^{-4\be}\left(\frac{\zeta_{\es_1}^2(\ell)\eta_{\es_1}(2\ell)}{
\es_2+\zeta_{\es_1}^2(\ell)}\tan^2(e^{-2\be}(\zeta_{\es_1}(\ell)(x+\al_1)+\eta_{\es_1}(\ell)(y+\be_1))+\ell_1)+2\eta_{\es_1}(2\ell)\right)
+\gm_1'\\-({\be'}'+2(\be')^2)(\es_1(x+\al_1)^2+(y+\be_1)^2)+\es_1{\al_1'}'x+{\be_1'}'y-\frac{\es_1(\al_1')^2
+(\be_1')^2}{2}.\hspace{1.7cm}(2.70)\end{eqnarray*}

\bibliographystyle{amsplain}

\begin{thebibliography}{10}



\bibitem{} M. J. Ablowitz and P. A. Clarkson, {\it Solitons,
Nonlinear Evolution Equations and Inverse Scattering}, London Math.
Soc Lect. Notes {\bf 149}, Cambridge University Press, 1991.

\bibitem{} D. Anker and N. C. Freeman, On the soliton solutions of the
Davey-Stewartson equation for long waves, {\it Proc. Roy. Soc.
London Ser. A} {\bf 360} (1978), 529-540.

\bibitem{} V. A. Arkadiev, A. K. Pogrebkov and M. C. Polivanov, Closed string
solution of the Davey-Stewartson equation, {\it Inverse Problem}
{\bf 5} (1989), L1-L6.

\bibitem{} V. A. Arkadiev, A. K. Pogrebkov and M. C. Polivanov, Inverse
sacttering transform method and soliton solution for the
Davey-Stewartson II equation, {\it Phys. D} {\bf 36} (1989),
189-197.

\bibitem{} P. A. Clarkson and S. Hood, New symmetry reductions and exact
solutions of the the Davey-Stewartson system. I. Reductions to
ordinary differential equations, {\it J. Math. Phys.} {\bf 35}
(1994), 255-283.

\bibitem{} A. Davey and K. Stewartson, On three-dimensional packets of
surface waves, {\it Proc. Roy. Soc. London Ser. A} {\bf 338} (1974),
101-110.

\bibitem{} J. T. Kirby and R. A. Dalrymple, Oblique envelope solutions of the
Davey-Stewartson equations in intermediate water depth, {\it Phys.
Fluids} {\bf 26} (1983), 2916-2918.

\bibitem{} C. R. Gilson and J. J. C. Nimmo, A direct method for dromion
solutions of the Davey-Stewartson equations and their asymptotic
properties, {\it Proc. Roy. Soc. London Ser. A} {\bf 435} (1991) ,
339-357.

\bibitem{} F. Guil and M. Manas, Deformation of the dromion and solutions of
the Davey-Stewartson I equation, {\it Phys. Lett. A} {\bf 209}
(1995), 39-47.

\bibitem{} M. Manas and P. Santini, Solutions of the Davey-Stewartson
equation with arbitrary rational localization and nontrivial
interaction, {\it Phsy. Lett. A} {\bf 227} (1997), 325-334.

\bibitem{} T. M. Malanyuk, Finite-gap solutions of the Davey-Stewartson 2
equations, {\it Russian Math. Surveys} {\bf 46} (1991), 193-194.

\bibitem{} T. M. Malanyuk, Finite-gap solutions of the Davey-Stewartson 1
equations, {\it Nonlinear Sci.} {\bf 4} (1994), 1-21.


\bibitem{} M. Omote, Infinite-dimensional symmetry algebras and an infinite
number of conserved quantities of the (2+1)-dimensional
Davey-Stewartson equation, {\it J. Math. Phys.} {\bf 29} (1988),
2599-2603.

\bibitem{} E. Saied, R. EI-Rahman and M. Ghonamy, On the exact
 solution of (2+1)-dimensional cubic nonlinear Schr\"{o}dinger
(NLS) equation", {\it J. Phy. A: Math. Gen.} {\bf 36} (2003),
6751-6770.

\bibitem{} J. van der Linden, Solutions of the Davey-Stewartson equation with
boundary condition, {\it Phys. Lett. A} {\bf 182} (1992), 155-189.


\bibitem{} Z. Wang and D. Guo, {\it Special functions}, World Scientific,
Singapore, 1998.

\bibitem{} X. Xu,  Quadratic-argument approach to nonlinear Schr\"{o}dinger
equation and coupled ones, {\it preprint}.


\end{thebibliography}

\end{document}